\documentclass[aps,reprint,prl,showpacs]{revtex4-1}

\usepackage[english]{babel}

\usepackage[margin=0.6in]{geometry}

\usepackage{mathtools}
\usepackage{amsmath}
\usepackage{amsfonts}
\usepackage{amssymb}
\DeclareMathOperator*{\rint}{\ThisStyle{\hstretch{1.3}{\rotatebox{18}
{$\SavedStyle\!\int\!$}}}}
\usepackage{scalerel}
\usepackage{graphicx}
\usepackage{bm}
\usepackage{afterpage}   

\usepackage{tikz,pgfplots}
\usetikzlibrary{matrix,arrows,decorations.markings}
\tikzset{->-/.style={decoration={markings,
  mark=at position .5 with {\arrow{>}}},postaction={decorate}},
  -<-/.style={decoration={markings,
    mark=at position .5 with {\arrow{<}}},postaction={decorate}}}

\newcommand{\rez}[1]{\frac{1}{#1}}

\usepackage{hyperref}
\usepackage[all]{hypcap} 

\def\beq{\begin{equation}}
\def\eeq{\end{equation}}
\def\bea{\begin{eqnarray}}
\def\eea{\end{eqnarray}}

\numberwithin{equation}{section}
\numberwithin{figure}{section}

\begin{document}

\title{Geometrically disordered network models, quenched quantum gravity,
and critical behavior at quantum Hall plateau transitions}

\author{I. A. Gruzberg}
\affiliation{Ohio State University, Department of Physics, 191 W. Woodruff
Ave, Columbus OH, 43210}

\author{A. Kl\"umper}
\affiliation{Bergische Universit\"at Wuppertal, Gau\ss stra\ss e 20, 42119
Wuppertal, Germany}

\author{W. Nuding}
\affiliation{Bergische Universit\"at Wuppertal, Gau\ss stra\ss e 20, 42119
Wuppertal, Germany}

\author{A. Sedrakyan}
\affiliation{Yerevan Physics Institute, Br. Alikhanian 2, Yerevan 36,
Armenia}

\begin{abstract}		
Recent high-precision results for the critical exponent of the localization
length at the integer quantum Hall (IQH) transition differ considerably
between experimental ($\nu_\text{exp} \approx 2.38$) and numerical
($\nu_\text{CC} \approx 2.6$) values obtained in simulations of the
Chalker-Coddington (CC) network model. We revisit the arguments leading to
the CC model and consider a more general network with geometric
(structural) disorder. Numerical simulations of this new model lead to the
value $\nu \approx 2.37$ in very close agreement with experiments. We argue
that in a continuum limit the geometrically disordered model maps to the
free Dirac fermion coupled to various random potentials (similar to the CC
model) but also to quenched two-dimensional quantum gravity. This explains
the possible reason for the considerable difference between critical
exponents for the CC model and the geometrically disordered model and may
shed more light on the analytical theory of the IQH transition. We extend
our results to network models in other symmetry classes.
\end{abstract}

\date{April 11, 2016}


\maketitle

\phantomsection
\setcounter{section}{1}
{\it Introduction.} The integer quantum Hall (IQH) transition
\cite{Huckestein-Scaling-1995} is the most prominent example of an Anderson
transition, a continuous quantum phase transition driven by disorder and
accompanied by universal critical phenomena \cite{Evers-Anderson-2008}.
Numerous experiments \cite{Wei-Experiments-1988, Koch-Experiments-1991,
Koch-Size-dependent-1991, Koch-Experimental-1992, Engel-Microwave-1993,
Wei-Current-1994} demonstrated scaling near the IQH transition characterized
by the localization length exponent $\nu$. The most recent and accurate
experimental value is $\nu_\text{exp} = 2.38 \pm 0.02$
\cite{Li-Scaling-2005, Li-Scaling-2009}. A similar value of $\nu$ was
observed at the IQH transition in graphene \cite{Giesbers-Scaling-2009},
confirming universality at the IQH transition.


The IQH effect is usually modeled by neglecting electron-electron
interactions, that is, within the paradigm of Anderson localization
\cite{Anderson-Absence-1958, Abrahams-Scaling-1979}. Existence of delocalized
states in disorder-broadened Landau levels, which is necessary to explain the
IQH transition, is consistent with the description of the transition by
a nonlinear sigma model with a topological term \cite{Levine-Electron-1983,
Weidenmuller-Single-1987}, and its two-parameter flow diagram
\cite{Kmelnitskii-Quantization-1983, Pruisken-Dilute-1985}. The critical
point of the sigma model should possess conformal invariance and be described
by a conformal field theory (CFT) with the central charge $c=0$
\cite{Gurarie-Conformal-2004}, due to the use of replicas or supersymmetry
(SUSY) to treat disorder averages. However, this fixed point is in the strong
coupling regime, and notable attempts at identifying the CFT
\cite{Zirnbauer-Conformal-1999, Bhaseen-Towards-2000, Tsvelik-Wave-2001,
Tsvelik-Evidence-2007} are inconclusive so far.

The IQH transition is related to the problem of disordered Dirac fermions
\cite{Ludwig-Integer-1994}. The generic model with random mass, scalar, and
gauge potentials is believed to have a fixed point in the universality class
of the IQH transition, but this fixed point is not perturbatively accessible.
A simplified model where only a random gauge potential is kept, is
analytically solvable, and the exact spectrum of multifractal (MF) exponents
describing the scaling of the moments of critical wave functions is known
\cite{Ludwig-Integer-1994, Chamon-Instability-1996,
Mudry-Two-dimensional-1996, Chamon-Localization-1996, Kogan-Liouville-1996,
Castillo-Exact-1997}.

More recently, alternative approaches to the IQH transition were advanced.
One is based on a mapping to a classical model and conformal restriction
\cite{Bettelheim-Quantum-2012}, and another uses symmetry properties of the
sigma model \cite{Gruzberg-Symmetries-2011, Gruzberg-Classification-2013,
Bondesan-Pure-2014} to derive exact symmetry properties of the MF spectra at
the IQH transition.
In spite of these successes, no theoretical predictions for the exponent
$\nu$ exist.

Much intuition about the IQH transition, as well as the most accurate
numerical estimates for critical exponents, come from a
the Chalker-Coddington (CC) network model \cite{Chalker-Percolation-1988,
Kramer-Random-2005}. The model is based on the semiclassical picture of
electrons drifting along the equipotential lines of a smooth disorder
potential. Tunneling across saddle points of the potential leads to
hybridization of the localized states and a possible delocalization.
In the CC model this picture is drastically simplified, and all scattering
nodes are placed at the vertices of a square lattice.
The CC model in various limits can be mapped both to the nonlinear sigma
model \cite{Read-1991, Zirnbauer-Towards-1994}, and the random Dirac fermions
\cite{Ho-Models-1996}.

The regular geometry of the CC model allows for an easy application of
numerical transfer matrix (TM) techniques \cite{MacKinnon-The-scaling-1983}.
The most recent and accurate implementations of this method
\cite{Slevin-Critical-2009, Obuse-Conformal-2010, Amado-Numerical-2011,
Obuse-Finite-2012, Slevin-Finite-2012, Nuding-Localization-2015}, as well as
other methods \cite{Dahlhaus-Quantum-2011, Fulga-Topological-2011} give the
value $\nu_\text{CC}$ in the range 2.56--2.62, which is definitely different
from the experimental value. One possible source for the discrepancy are
electron-electron interactions whose effect on the scaling near the IQH
transition has been studied in Refs. \cite{Lee-Effects-1996,
Wang-Short-range-2000, Burmistrov-Wave-2011}. It was shown there that
short-range interactions are irrelevant at the IQH critical point, and should
not modify the value of $\nu$. This leaves the option that the Coulomb
interaction may play a dominant role in experimental systems, but this issue
is not fully understood, and remains unresolved.

Here we propose another possible explanation for why the value of
$\nu_\text{CC}$ differs from $\nu_\text{exp}$, namely that the CC model does
not capture all types of disorder that are relevant at the IQH transition.
Indeed, saddle points that connect the ``puddles'' of filled electron states
do not form a regular lattice, and around each ``puddle'' there may be any
number of them. Taking this into account leads us to consider structurally
disordered, or {\it random networks} (RNs) that better represent the physics
in a smooth disorder potential and strong magnetic field.

\begin{figure}[t]
\centering
\includegraphics[height=3.7cm,bb=1 1 508 474]{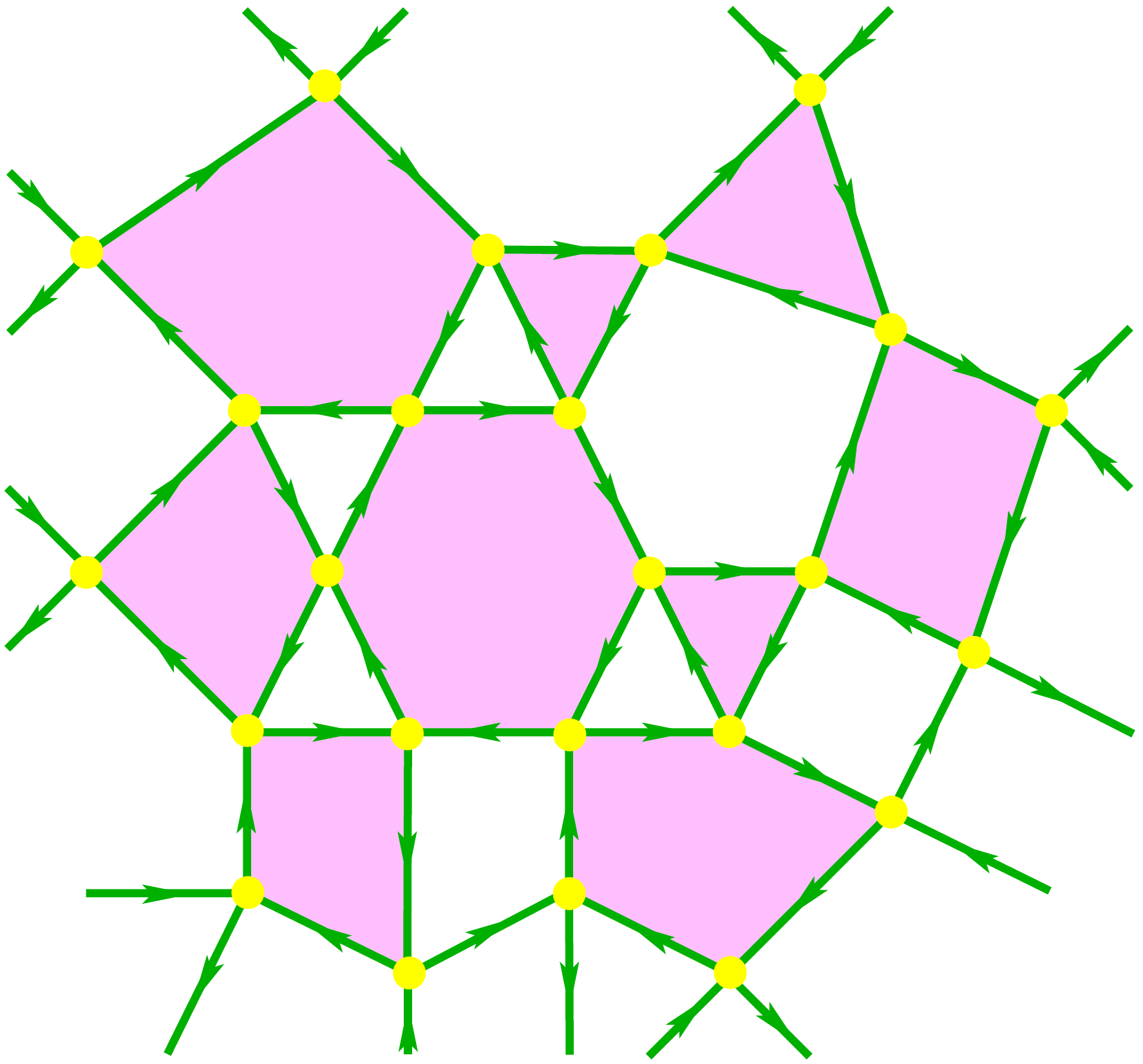}
\hfil
\includegraphics[height=3.7cm,
angle=0]{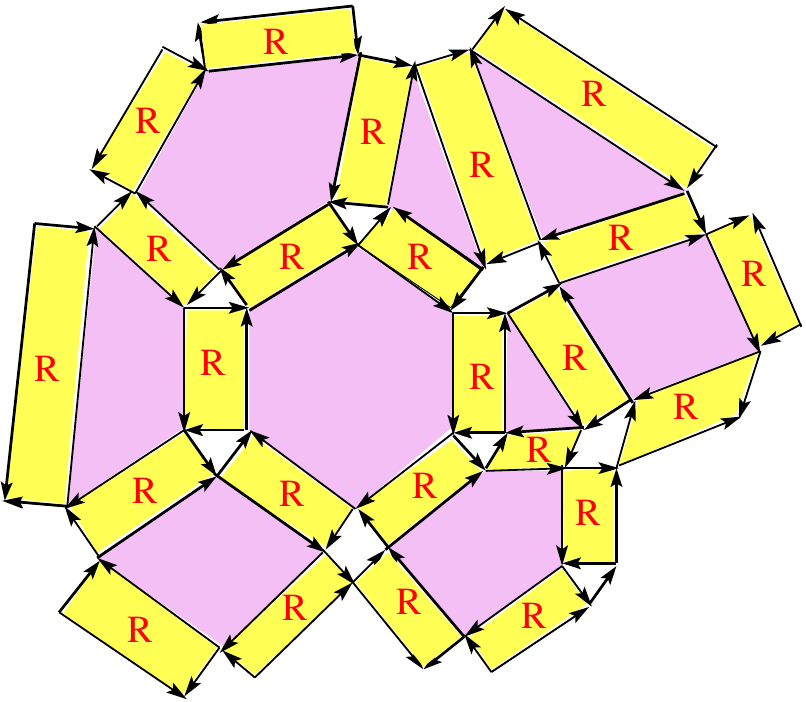}
\caption{Left: a random graph. Right: the corresponding random Manhattan
lattice.}
\label{fig:random-graph-and-medial}
\vskip -5mm
\end{figure}

Let us list the main results of this paper. (1) We argue that an ensemble of
RNs can be mapped in a continuum limit to the problem of free Dirac fermions
coupled to random potentials (similar to the CC model) and also to
two-dimensional quantum gravity (2DQG). Coupling to 2DQG modifies critical
exponents of statistical mechanics models \cite{Knizhnik-Fractal-1988,
David-Conformal-1988, Distler-Conformal-1988, Kazakov-Exactly-1988,
Kazakov-Recent-1988, Kazakov-Percolation-1989, Duplantier-Geometrical-1990}.
We suggest that a similar modification happens for RNs. (2) We demonstrate
that RNs can be effectively constructed starting with the CC network and
appropriately modifying it. The modified RNs can be numerically simulated,
and for certain values of parameters specifying the geometric disorder, we
obtain the localization length exponent
$\nu = 2.374 \pm 0.018$,
in excellent
agreement with experiments. (3) We extend these ideas to
quantum Hall
transitions in symmetry classes C and D in the classification of Refs.
\cite{Zirnbauer-Riemannian-1996, Altland-Nonstandard-1997}. Properties of
these transitions map to classical statistical mechanics models which were
studied on random lattices, and for which the shift in critical exponents is
given by the KPZ relation \cite{Knizhnik-Fractal-1988, David-Conformal-1988,
Distler-Conformal-1988} from the theory of 2DQG. This fact allows us to
predict various exact critical exponents for these transitions.

{\it Random networks.} The network models we consider are built on planar
directed graphs where every vertex has two incoming and two outgoing edges.
The in- and out- edges, also called links of the network, alternate as one
goes around a vertex (a node). Such graphs divide the plane into two sets of
polygonal faces with opposite orientations of their edges, see Fig.
\ref{fig:random-graph-and-medial}, left. We will only consider connected
graphs, which are exactly the Feynman graphs of zero-dimensional (complex)
matrix $\phi^4$ theory in the planar (large $N$) limit
\cite{'tHooft-Planar-1973, Brezin-Planar-1977}.

\begin{figure}[t]
\centering
\includegraphics[
width=0.44\columnwidth,
angle=0]{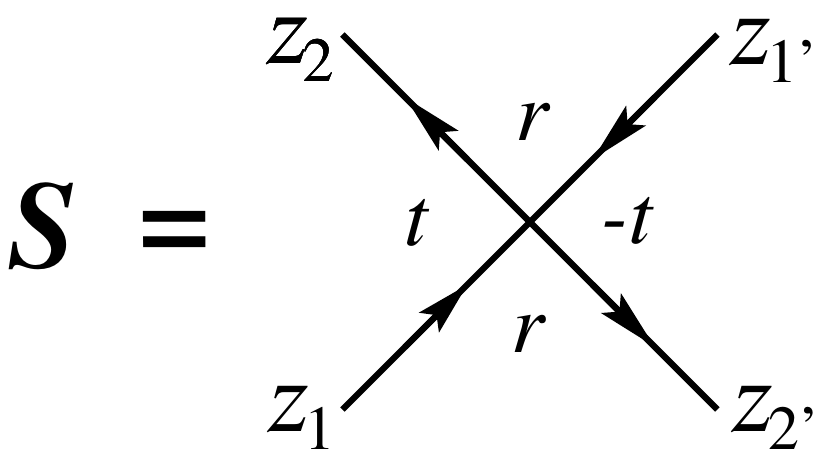}
\hfil
\includegraphics[
width=0.46\columnwidth,
angle=0]{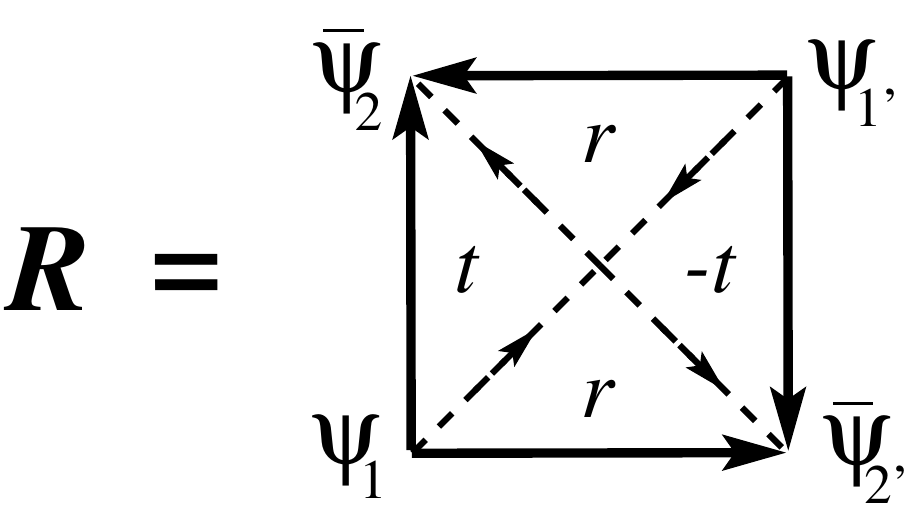}
\caption{Left: an $S$ matrix. Right: the corresponding $R$ matrix.}
\label{fig:S-and-R}
\vskip -5mm
\end{figure}

A state of the network model on a given random graph is represented by a
complex vector $Z \in \mathbb{C}^N$, where $N$ is the number of edges of the
graph, and each component $z_e$ corresponds to the complex flux on the edge
$e$. The model includes random scattering matrices connecting incoming $z_1,
z_{1'}$ and outgoing $z_2, z_{2'}$ fluxes (see Fig. \ref{fig:S-and-R}, left):
\begin{align}
\Big(\!\begin{array}{c} z_2 \\ z_{2'} \end{array} \!\!\Big)
= {\cal S} \Big(\!\begin{array}{c} z_1 \\ z_{1'} \end{array}\!\! \Big)
= \Big(\!\begin{array}{cc} t e^{i\gamma} & r e^{i\gamma'} \\
r e^{i\gamma} & -t e^{i\gamma'}\end{array} \!\!\Big)
\Big(\!\begin{array}{c} z_1 \\ z_{1'} \end{array} \!\!\Big),
\end{align}
placed at the vertices. The scattering amplitudes satisfy $t^2 + r^2 =1$, and
the scattering phases $\gamma$, $\gamma'$ are random.

Evolution of the states of the network in discrete time steps is specified by
an $N\times N$ unitary matrix $U$ composed of all node scattering matrices
\cite{Klesse-Universal-1995}. In this description the basic object
is the resolvent $(1 - e^{-\eta} U)^{-1}$. Its matrix element (a Green
function) can be written as a superintegral
\begin{align}
\label{GF}
G(e_1, e_2; \eta) &= \!\!\! \rint \!\!\! {\cal D} \Psi \, \psi_{e_1}
{\bar \psi}_{e_2} e^{-\sum_{e,e'} {\bar \Psi}_{e} (1 - e^{-\eta}U)_{ee'}
\Psi_{e'}}
\end{align}
where $e_1$, etc., label edges of the graph, and ${\bar \Psi}_{e} =
({\bar\phi}_{e}, {\bar\psi}_{e})$ is a supervector assigned to the edge $e$,
see Refs. \cite{Janssen-Point-contact-1999, Cardy-Network-2005} for details.
The real part of the parameter $\eta$ plays the role of the imaginary part of
the energy (level broadening) in the Hamiltonian description. For our
purposes it is sufficient to take $\eta=0$ in what follows.

Formulation of a random network as a lattice model appeared in Ref.
\cite{Sedrakyan-3DIM-1987} in connection with the so called sign factor
problem in the string representation of the 3D Ising model. This approach was
further developed in Refs. \cite{Sedrakyan-Edge-1999,
Sedrakyan-Integrable-2002, Sedrakyan-Action-2003, Khachatryan-Grassmann-2009,
Khachatryan-Network-2010}. Following these references, we connect the
midpoint of each edge $e$ ``forward'' to two other midpoints by two vectors
$\xi_e$. Then a scattering node is replaced by a rectangle (see Fig.
\ref{fig:S-and-R}, right), and we get an alternative representation of the RN
as a random Manhattan lattice (ML), see the right part of Fig.
\ref{fig:random-graph-and-medial}.
The action for the RN written as
\begin{align}
\label{action}
S = \sum_e {\bar \Psi}_{e} \Psi_e - \sum_{e,\xi_e} t_{e,\xi_e} e^{i\gamma_e}
{\bar \Psi}_{e+\xi_e} \Psi_e
\end{align}
represents hopping of fermions and bosons on the random ML, and the hopping
amplitudes take values $r$ and $\pm t$ depending on the vector $\xi_e$.

The SUSY method of Refs. \cite{Janssen-Point-contact-1999,
Cardy-Network-2005} is designed to describe only single-particle problems,
while the approach of Refs. \cite{Sedrakyan-Edge-1999,
Sedrakyan-Integrable-2002, Sedrakyan-Action-2003, Khachatryan-Grassmann-2009,
Khachatryan-Network-2010} allows to consider interacting particles. To this
end one uses the second quantization, and the scattering matrices at the
nodes are ``promoted'' to R-matrices acting in the tensor product of Fock
spaces attached to edges of the network (see Fig. \ref{fig:S-and-R}, right).
On a random ML the R-matrices are represented by the quadrangular faces
surrounding the scattering nodes, see Fig. \ref{fig:random-graph-and-medial}.
The trace of the product of the R-matrices over all nodes of the network
gives the partition function. For a general interacting case the SUSY method
does not apply, and one has to use replicas to treat disorder. In this paper
we do not include interactions and continue to use SUSY. Then writing the
trace of the product of the R-matrices in the basis of (super-)coherent
states for each of the (super-)Fock spaces on the edges, we obtain the same
action (\ref{action}).

{\it Continuum limits.} For the regular CC model the ML is a square lattice
with vertices labeled by the Cartesian coordinates~$x^\mu$ ($\mu = 1,2$). The
vectors $\xi_e$ are $\pm \epsilon {\hat x}_\mu$, where ${\hat x}_\mu$ are
unit vectors, and $\epsilon$ is the lattice spacing. Near the critical point
of the CC model ($t_c = r_c = 1/\sqrt{2}$) the variations of the phases
$\gamma_e$ and the fields $\Psi_e$ are slow, and we can pass to a continuum
limit by expanding $\Psi_{x + \epsilon{\hat x}_\mu} \approx (1 + \epsilon
\partial_\mu)\Psi_x$ and rescaling the fields $\Psi(x)$ in the continuum. In
the limit we obtain, as in Ref. \cite{Ho-Models-1996}, the action of the
Dirac fermions (and their bosonic partners)
\begin{align}
\label{S2}
S = \!\! \rint \!\!\! d^2 x \, \bar\Psi\big[\sigma^\mu
\big( i \overset{\text{\tiny$\leftrightarrow$}}{\partial}_\mu
+ A_\mu \big) + m \sigma^3 + V \big]\Psi,
\end{align}
where $\overset{\text{\tiny$\leftrightarrow$}}{\partial}_\mu =
(\overset{\text{\tiny$\rightarrow$}}{\partial}_\mu -
\overset{\text{\tiny$\leftarrow$}}{\partial}_\mu)/2$, the mass $m \propto
r-r_c$, and the (random) gauge $A_\mu(x)$ and scalar $V(x)$ potentials arise
as certain combinations of the random phases $e^{i\gamma_e}$.

\begin{figure}[t]
\centering
\includegraphics[
width=0.7\columnwidth,
angle=0]{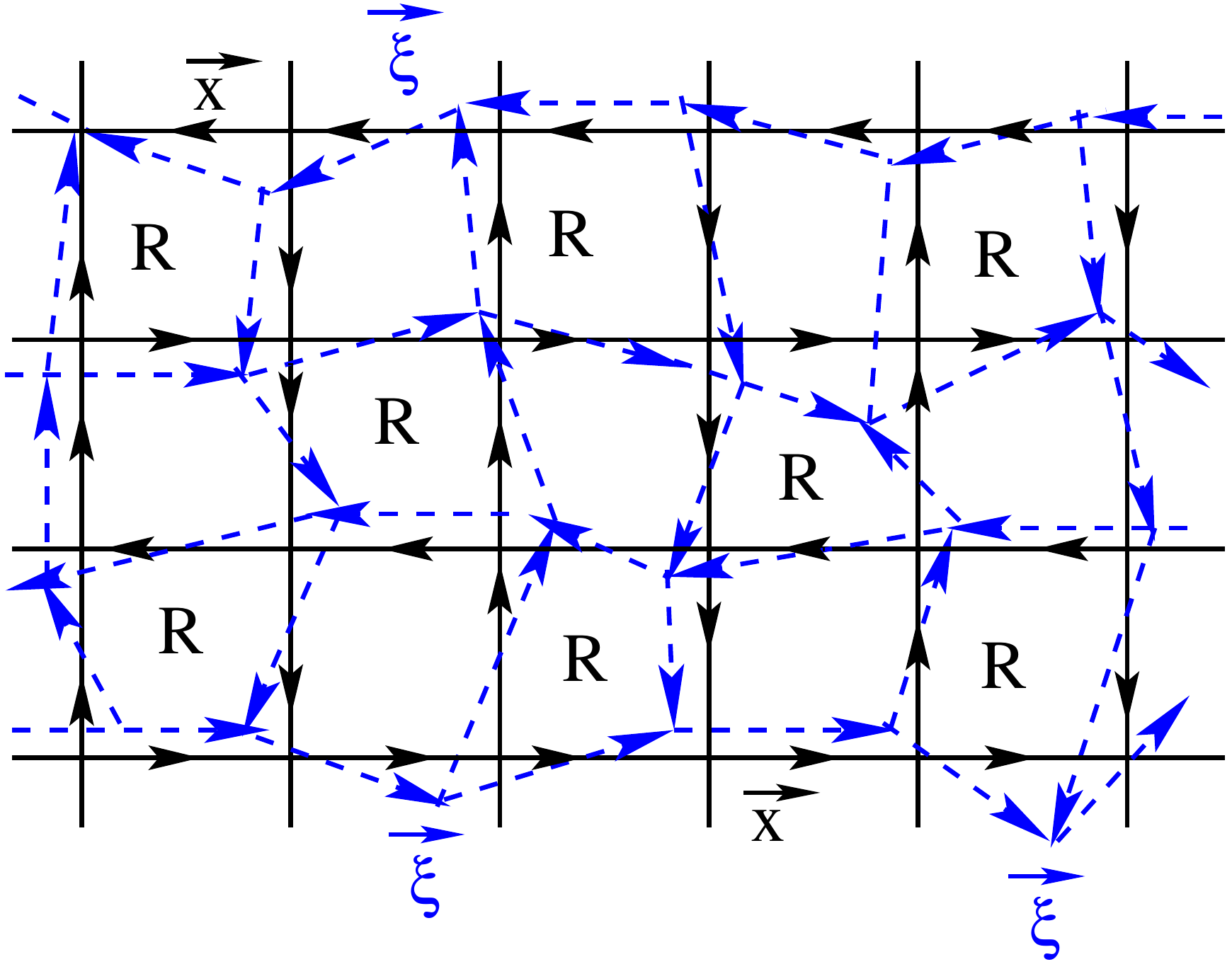}
\vskip -2mm
\caption{Weakly random Manhattan lattice.}
\label{fig:flat-random-network}
\vskip -4mm
\end{figure}

Let us now consider the random ML shown in Fig.
\ref{fig:flat-random-network}. This lattice is not very different from the
regular square lattice, its faces are still quadrangles, and we can introduce
(curvilinear) coordinates $\xi^a$ ($a=1,2$) following the vectors $\xi_e$ in
a natural way. It is clear that the physics cannot depend on the choice of
coordinates, so we can use either $\xi^a$ or $x^\mu$ coordinates. We can use
the formalism of frames (vielbeins) of differential geometry
\cite{Nakahara-Geometry-2003} to relate coordinate and orthonormal bases of
vectors ${\hat x}_\mu = e_\mu^a \partial/\partial \xi^a$ and forms $dx^\mu =
e_a^\mu d\xi^a$, as well as the volume elements $d^2 x = e \, d^2 \xi$, where
$e= \det e_a^\mu$.
The action (\ref{S2}) written in arbitrary coordinates and invariant under
coordinate changes becomes
\begin{align}
\label{S3}
S = \!\! \rint \!\!\! d^2 \xi \, e \, \bar\Psi \big[\sigma^\mu e_\mu^a
\big(i \overset{\text{\tiny$\leftrightarrow$}}{\partial}_a
+ A_a \big) + m \sigma^3 + V \big]\Psi.
\end{align}

The action (\ref{S3}) is that of 2D fermions interacting with random gauge
and scalar potentials as well as random geometry (gravity). In the case of
weakly deformed lattices, Eqs. (\ref{S2}) and (\ref{S3}) are equivalent, they
both describe the system on a flat surface. We propose that random frames can
account for more complicated situations that correspond to curved surfaces
represented by random graphs. In this case we define frames locally, on a
given coordinate chart, and then connect them on overlapping charts by
transition functions. The result is still given by Eq. (\ref{S3}), but now we
are supposed to average over ``arbitrary'' frame configurations. The above
arguments leave open the question of the functional measure on random
surfaces. We believe that the requirements of diffeomorphism and conformal
invariance determine the appropriate measure uniquely, the same way it is
fixed in string theory \cite{Polyakov-Quantum-1981}.

The need to average observables over random geometry means that our system is
coupled to {\it quenched} quantum gravity. However, in the SUSY formalism the
partition function of a disordered system is always unity (implying $c=0$ for
the CFT of the critical point), and there is no difference between quenched
and annealed gravity.

It is known that 2DQG modifies critical exponents of a CFT placed on a
fluctuating surface in the way given by the KPZ relation
\cite{Knizhnik-Fractal-1988, David-Conformal-1988, Distler-Conformal-1988}.
The relation has been verified by solutions of critical models of statistical
mechanics (related to the so-called minimal CFTs
\cite{Belavin-Infinite-1984}) defined on random graphs
\cite{Kazakov-Exactly-1988, Kazakov-Recent-1988, Kazakov-Percolation-1989,
Duplantier-Geometrical-1990}. When $c=0$, as for Anderson transitions and
critical percolation,  the relation is
\begin{align}
\label{KPZ}
\Delta = (\sqrt{1 + 24 \Delta_0} - 1)/4,
\end{align}
where $\Delta_0$ ($\Delta$) are chiral dimensions of operators on a flat
(fluctuating) surface. Whether this relation can explain the difference
between $\nu_\text{CC}$ and $\nu_\text{exp}$ is to be seen. However, Eq.
(\ref{KPZ}) should be applicable to properly defined MF exponents of critical
wave functions at the IQH transition, as well as other 2D Anderson
transitions.

\begin{figure}[t]
\centering
\includegraphics[
width=0.8\columnwidth,
angle=0]{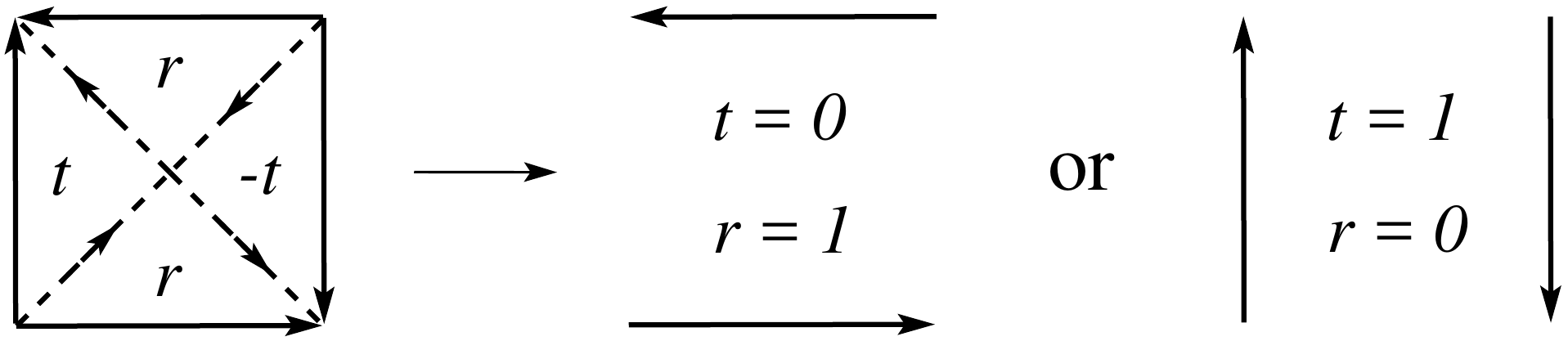}
\vskip 2mm
\includegraphics[
width=0.95\columnwidth,
angle=0]{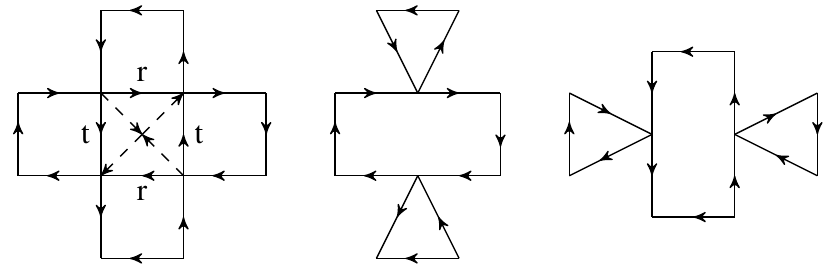}
\caption{Top: opening a node. Bottom: The resulting modifications of the
CC network.}
\label{fig:open-nodes}
\vskip -4mm
\end{figure}

{\it Construction and simulation of RNs.} To simulate RNs numerically, we
adopt the following construction. Starting with the regular CC network, at
each node we set $t=0$ with probability $p_0$,  $t=1$ with probability $p_1$,
and leave the node unchanged with probability $p_c = 1 - p_0 - p_1$. The
modified nodes with $t=0$ ($t=1$) are ``open'' in the horizontal (vertical)
direction, and opening a node changes the four adjacent square faces into two
triangles and one hexagon, see Fig. \ref{fig:open-nodes}. Repeated opening of
nodes can produce tilings of the plane by polygons with arbitrary numbers of
edges. At the same time, our construction still allows us to use the transfer
matrix (TM) of the CC model, but with modified $t$ and $r$ amplitudes.

To maintain statistical isotropy of the model, we choose $p_0 = p_1$. In this
case we expect that the critical point is still given by the value
$t_c^2=1/2$ for the unchanged nodes. Moreover, in this paper we fix $p_0 =
p_1 = p_c = 1/3$.

We simulate the modified networks on strips of different width $M$ (the
number of nodes per column) varying from $20$ to $200$, the length $L=5\cdot
10^6$, and a range of the parameter $x$ which encodes deviations of $t$ from
$t_c$ \footnote{See Supplementary material.}. We use the LU decomposition of
TMs \cite{numerical_recipes}. Since $t$ and $r$ appear in the denominators of
the matrix elements of TMs, making them zero is a singular procedure, related
to the disappearance of two horizontal channels upon opening a node in the
vertical direction. To overcome this difficulty, for every open node we take
either $t$ or $r$ to be equal to $\varepsilon \ll 1$. We then look at how the
resulting Lyapunov exponents depend on $\epsilon$. We found that the results
saturate at $\varepsilon = 10^{-5}$, and there are no changes when reducing
$\epsilon$ to $10^{-7}$. For even smaller $\varepsilon$ the results start
changing again. This is to be expected because the large differences of
values in the entries of TMs cause numerical instabilities for the LU
decomposition. We have chosen $\varepsilon=10^{-6}$ for our calculations.

The smallest Lyapunov exponent $\gamma$ is expected to have the following
finite-size scaling behavior:
\begin{equation} \label{ren_equ}
	\gamma M=\Gamma[M^{1/\nu}u_0(x), M^y u_1(x)].
\end{equation}
Here $u_0(x)$ is the relevant field and $u_1(x)$ the leading irrelevant
field. The relevant field vanishes at the critical point, and $y<0$. The
fitting procedure of our numerical results, as well as the error analysis are
presented in the Supplementary material. The results of the analysis are
\begin{align}
\nu &= 2.374 \pm 0.018, & y = -0.35 \pm 0.05.
\end{align}
This value of $\nu$ is surprisingly close to $\nu_\text{exp}$, which suggests
that the structural disorder is, indeed, a relevant perturbation that
modifies the critical behavior.

{\it Other symmetry classes.} Network models can be constructed for all 10
symmetry classes of disordered systems identified in Refs.
\cite{Zirnbauer-Riemannian-1996, Altland-Nonstandard-1997}. Superconductors
with broken time-reversal invariance in 2D can exhibit QH transitions where
the spin (class C) \cite{Kagalovsky-Quantum-1999, Senthil-Spin-1999} and
thermal (class D) \cite{Senthil-Quasiparticle-2000} conductivities jump in
quantized units. The ideas developed above apply to network models for these
transitions. In addition, both SQH and TQH are simpler than the IQH since
many of their properties can be determined from mappings to classical models.

The regular network in class C was mapped to classical bond percolation on a
square lattice \cite{Gruzberg-Exact-1999, Beamond-Quantum-2002,
Mirlin-Wavefunction-2003}. Many exact results are known for classical
percolation. Thus, the mapping has lead to a host of exact critical
properties at the SQH transition \cite{Gruzberg-Exact-1999,
Mirlin-Wavefunction-2003, Cardy-Linking-2000, Subramaniam-Surface-2006,
Subramaniam-Boundary-2008, Bondesan-Exact-2012, Bhardwaj-Relevant-2015}. The
mapping was extended to network models in class C on arbitrary graphs
\cite{Cardy-Network-2005}.

\begin{figure}[t]
\centering
\includegraphics[height=3.7cm,
angle=0]{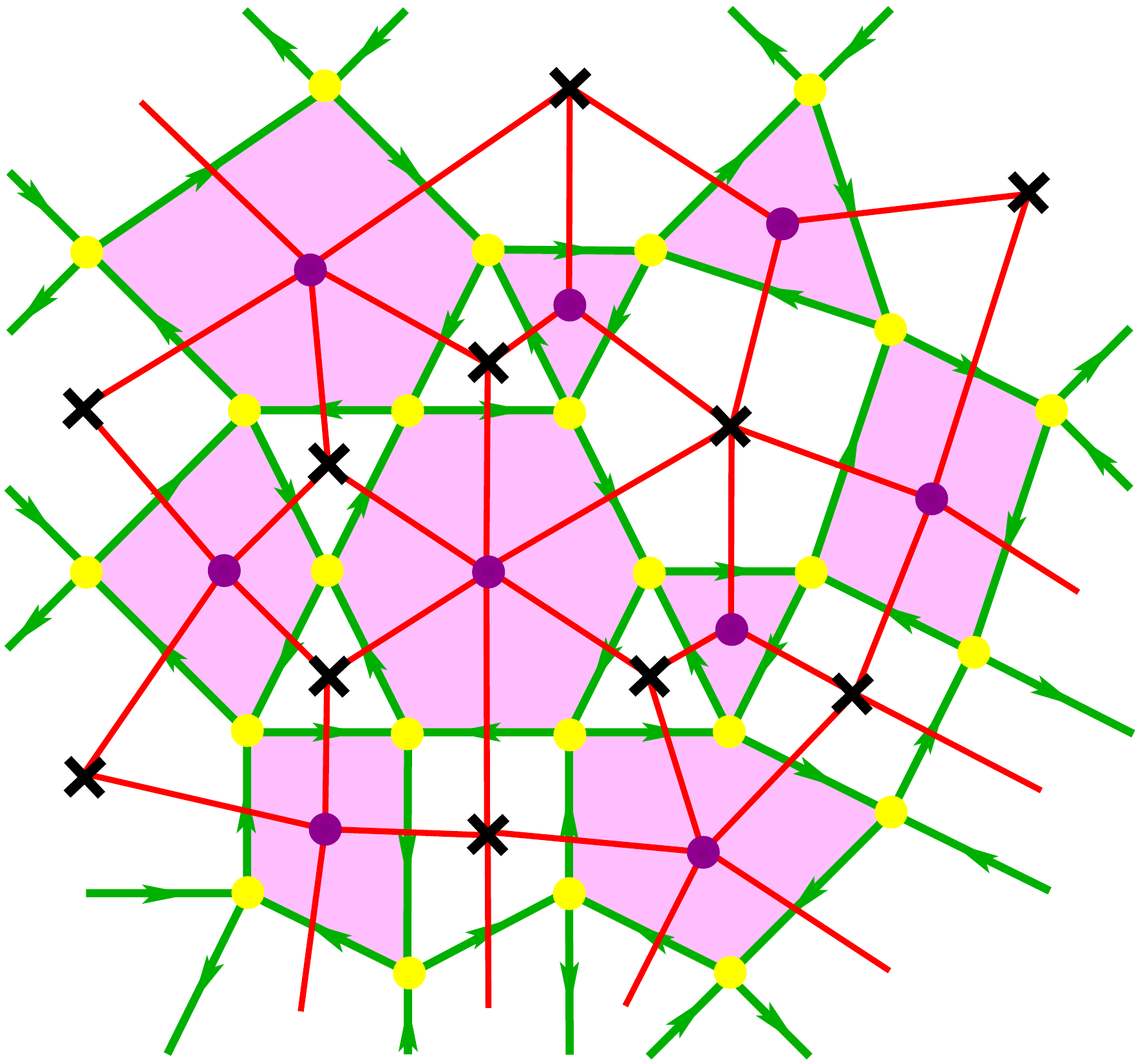}
\hfil
\includegraphics[height=3.7cm,
angle=0]{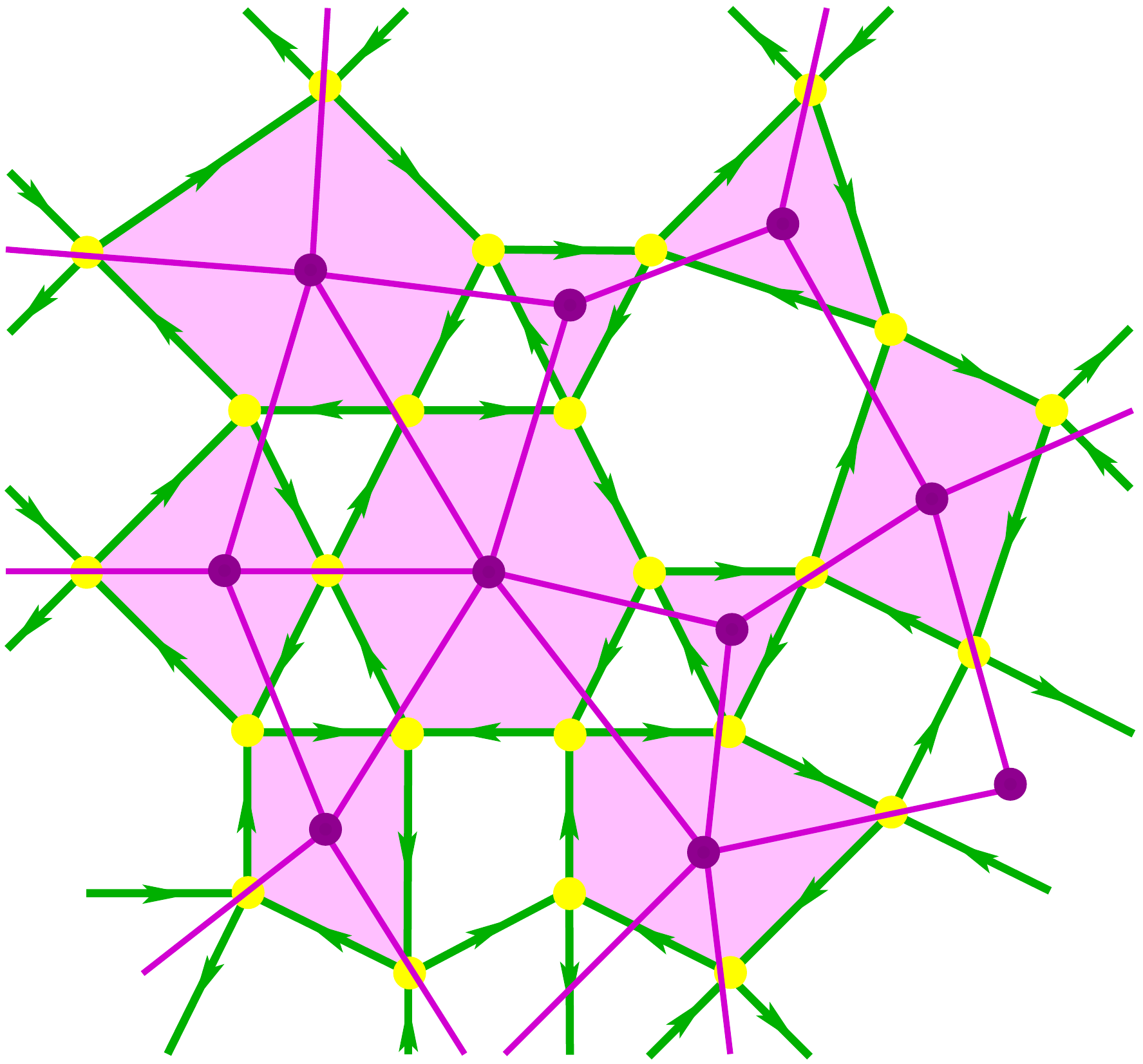}
\caption{Left: original RN and its dual. Right: percolation lattice.}
\label{fig:random-graph-and-dual}
\vskip -5mm
\end{figure}

The graphs relevant for our study are shown in Fig.
\ref{fig:random-graph-and-dual}. For a given RN we draw the dual bipartite
graph with dots on the shaded faces and crosses on the empty faces of the
original RN. The dual graph forms a random quadrangulation of the plane. We
now dissect all quadrangles by diagonals connecting the dots, and remove the
crosses and all edges connected to them. This results in a lattice (Fig.
\ref{fig:random-graph-and-dual}, right) on which the classical bond
percolation should be considered.



Critical bond percolation on random quadrangulations (or their duals) was
considered in Ref. \cite{Kazakov-Percolation-1989}, and it was shown that the
KPZ relation (\ref{KPZ}) is valid in this case. We believe that the SQH
transition on RNs lies in the same universality class, and that Eq.
(\ref{KPZ}) can be applied to all critical exponents obtained in Refs.
\cite{Gruzberg-Exact-1999, Mirlin-Wavefunction-2003,
Subramaniam-Surface-2006, Subramaniam-Boundary-2008, Bondesan-Exact-2012,
Bhardwaj-Relevant-2015}. This includes, in particular, the dimension of the
``two-leg'' operator that determines the localization length exponent $\nu$
as well as a few MF exponents.

The TQH transition in class D can also be described and simulated by a
network model \cite{Cho-Criticality-1997, Chalker-Thermal-2002,
Merz-Two-dimensional-2002, Mildenberger-Density-2007}. Its effective field
theory (without geometric disorder) is given by the Majorana fermions with
random mass, the same theory that describes the critical Ising model with a
weak bond disorder \cite{Senthil-Quasiparticle-2000,
Bocquet-Disordered-2000}. The random mass is a marginally irrelevant
perturbation, and critical exponents at the transition are given by their
Ising model values. When the model is coupled to 2DQG, we still should
consider the quenched situation, and the critical exponents should be
modified according to Eq. (\ref{KPZ}), see \cite{Janke-Two-dimensional-2006}
and references therein.

{\it Discussion and outlook.} The geometric disorder that we simulate by a
modified CC model can be viewed as randomness in the heights $V$ of the
saddle points in the disorder potential. Indeed, it is known that (at zero
energy) $t^2 = (1 + e^{-V})^{-1}$ \cite{Fertig-Transmission-1987}. Our choice
of $t$ is described by the tri-modal distribution
$P(V) = p_0 \delta(V - 2 \ln \varepsilon) + p_c \delta(V - V_0) + p_0
\delta(V + 2 \ln \varepsilon).$
Previous studies of random $V$ \cite{Lee-Quantum-1993,
Evers-Semiclassical-1998} focused instead on the uniform distribution in the
interval $V \in [-W, W]$ or the bimodal distribution $P(V) = [\delta(V-W) +
\delta(V+W)]/2$. No choice of $W$ gives our type of randomness when $p_c >
0$. However, for $p_c = 0$ our distribution becomes bimodal, and describes
classical percolation with $\nu = 4/3$. The other extreme, $p_c = 1$, gives
the regular CC model. Since we only simulated the point $p_c = 1/3$, we
cannot distinguish the following three possibilities: 1) a novel fixed point
at a finite $p_c$, 2) a crossover from percolation to CC criticality, 3) a
line of fixed points. We plan to study other values of $p_c$ to determine
which scenario is actually realized.

We also plan to simulate RNs in classes C and D, and try to solve the classical
percolation problem on relevant graphs using matrix models techniques. We
will, furthermore, consider the problem of Dirac fermions in an Abelian
random gauge potential coupled to 2DQG, and determine the MF spectrum of the
wave functions in order to test the applicability of the KPZ relation
(\ref{KPZ}).

{\it In summary}, we have considered the possibility that a certain type of
geometric (structural) disorder, previously missed in the study of the IQH
transition, may change the universality class. Our numerical simulations
support this idea. We have also proposed that the proper framework for a
field-theoretic description of this type of disorder is provided by 2DQG
coupled to matter fields. These ideas can be applied to other 2D Anderson
transitions.

\begin{acknowledgments}
A.~S. thanks the Theoretical Physics group at Wuppertal University for
hospitality. A.~S.~and A.~K.~acknowledge support by DFG grant KL 645/7-1.
A.~S.\ was partially supported by ARC grant 15T-1C058. I.~G. was partially
supported by the NSF Grant No. DMR-1508255. We are grateful to R.~A.~Roemer
and A.~W.~W. Ludwig for helpful discussions. Extensive calculations have been
performed on Rzcluster (Aachen), PC\textsuperscript{2} Paderborn) and
particularly on JUROPA (J\"ulich). The authors gratefully acknowledge the
computing time granted by the John von Neumann Institute for Computing (NIC)
and provided on the supercomputer JUROPA at J\"ulich Supercomputing Centre
(JSC).
\end{acknowledgments}


%

\clearpage

\onecolumngrid
\section*{Supplemental material \\
    Network model for plateau transitions in the quantum Hall effect
  }
  \setcounter{section}{2}
  \setcounter{figure}{0}
  \setcounter{equation}{0}
\twocolumngrid

We calculate numerically the localization length index $\nu$ in the
Chalker-Coddington (CC) network suitably modified to represent a random network. We use one relevant field and one irrelevant field in the fitting procedure. The results lead to the value $\nu \approx 2.37$ for the modified model, in very close agreement with experiments.

\subsection{Model description}
\label{model_desc}

For the calculation of critical indices we used the transfer-matrix method
developed in \cite{mackinnon1981scaling, mackinnon1983scaling}. To
calculate the smallest Lyapunov exponent of the CC-model it is
necessary to calculate a product $T_L=\prod_{j=1}^L M_1 U_{1j}M_2 U_{2j}$ of layers of transfer matrices $M_1 U_{1j}M_2 U_{2j}$ corresponding to two
columns $M_1$ and $M_2$ of vertical sequences of 2x2 scattering nodes,
\begin{equation} \label{M1}
   M_1= \begin{tikzpicture}[baseline=(current bounding box.center), ultra thick, loosely dotted]
				\matrix(M1)[matrix of math nodes,
										nodes in empty cells,
										right delimiter={)},
										left delimiter={(}]
				{
					B^1 & 0   & 	& 0  	\\
					0   &	B^1	&  	& 		\\
							& 		& 	& 0 	\\
					0 	& 		& 0	& B^1	\\
				};
				\draw (M1-2-2)--(M1-4-4);
				\draw (M1-1-2)--(M1-1-4);
				\draw (M1-4-1)--(M1-4-3);
				\draw (M1-2-1)--(M1-4-1);
				\draw (M1-1-4)--(M1-3-4);
				\draw (M1-1-2)--(M1-3-4);
				\draw (M1-2-1)--(M1-4-3);
			\end{tikzpicture}
\end{equation}
and
\begin{equation}\label{M2}
	M_2=\begin{tikzpicture}[baseline=(current bounding box.center), ultra thick, loosely dotted]
				\matrix (M2) [matrix of math nodes,nodes in empty cells,right delimiter={)},left delimiter={(}] {
					B^2_{22}	&	0  	&		&	0	&	B^2_{21}		\\
					0					&	B^2	&		&			&	0					\\
					 					& 		&		&			&						\\
					0      		&			&		&	B^2	&	0					\\
					B^2_{12}	&	0		&		& 0		&	B^2_{11}	\\
				};
				\draw (M2-1-2)--(M2-1-4);
				\draw (M2-1-2)--(M2-4-5);
				\draw (M2-2-1)--(M2-4-1);
				\draw (M2-2-1)--(M2-5-4);
				\draw (M2-2-2)--(M2-4-4);
				\draw (M2-2-5)--(M2-4-5);
				\draw (M2-5-2)--(M2-5-4);
			\end{tikzpicture}
\end{equation}
with
\begin{equation}
	B^1=\begin{pmatrix}
				1/t & r/t \\
				r/t & 1/t
			\end{pmatrix}
	\qquad \text{and} \qquad
	B^2=\begin{pmatrix}
				1/r & t/r \\
 				t/r & 1/r
			\end{pmatrix}
\end{equation}
The $U$-matrices have a simple diagonal form with independent phase factors $U_{nm}=\exp{(i\alpha_n)}\,\delta_{nm}$ for $U=U_{1j}$ and $U_{2j}$. Here $t$ and $r$ are the transmission and reflection amplitudes at each node of the regular lattice which are parameterized by
\begin{equation} \label{rt}
  t=\rez{\sqrt{1+e^{2x}}} \qquad \text{and} \qquad r=\rez{\sqrt{1+e^{-2x}}}.
\end{equation}
The parameter $x$ corresponds to the Fermi energy measured from the
Landau band center scaled by the Landau band width (with the critical point at $x=0$). The phases $\alpha_{n}$ are random variables uniformly distributed in the range $[0,2\pi)$, reflecting that the phase of an electron approaching a saddle point of the random potential is arbitrary.

To simulate random networks (RNs) numerically, we remove scattering nodes by opening them in horizontal or vertical direction with probabilities $p_0$ and $p_1$ by adopting the following construction. Starting with the regular CC network, at each node we set $t=\varepsilon\ll 1$ with probability $p_0$, $t=\sqrt{1-\varepsilon^2}$ with probability $p_1$, and leave the node unchanged with probability $p_c = 1 - p_0 - p_1$. Here the small number $\varepsilon$ is chosen as $\varepsilon=10^{-6}$: We found that the results saturate already at $\varepsilon = 10^{-5}$, and there are no changes when reducing $\varepsilon$ to $10^{-7}$. For even smaller $\varepsilon$ the results start changing again due to precision issues of the numerics.

Furthermore, in this report we use $p_0=p_1=p_c=1/3$.

\subsection{The fitting procedure}
For the scaling behavior of the Lyapunov exponent $\gamma$ near the critical
point we expect the finite size dependence
\begin{equation} \label{eq:ren_equ}
	\gamma\cdot M=\Gamma(M^{1/\nu}u_0,M^y\,u_1) ,
\end{equation}
Here we have taken into account the relevant field with exponent $\nu$ and the leading irrelevant field with exponent $y$. $M$ is the number of $2 \times 2$ blocks in the transfer matrices ($=$ half the number of horizontal channels of the lattice), $u_0=u_0(x)$ is the relevant field and $u_1=u_1(x)$ the leading irrelevant field. It is known that the relevant field vanishes at the critical point, and that $y<0$.

On the left hand side of Eq. \eqref{eq:ren_equ} we use the numerical results for the eigenvalues of $T_L$, where we are particularly interested in the eigenvalue closest to 1. The Lyapunov exponent $\gamma$ is the smallest
positive eigenvalue of
\bea
\label{LE}
\lim_{L\to\infty}\frac{\log[T_L T_L^\dagger]}{2L}\ ,
\eea
which we calculate for various combinations of the parameter $x$ and the lattice width $M$. The right hand side of \eqref{eq:ren_equ} is expanded in a series in $x$ and powers of $M$, and the expansion coefficients are obtained from a fit. Some coefficients in this expansion vanish due to a symmetry argument \cite{SlevinOhtsuki2009}. If $x$ is replaced by $-x$ we see from \eqref{rt} that $t$ turns into $r$ and vice versa. Due to the periodic boundary conditions the lattice is unchanged. Therefore the left hand side of \eqref{eq:ren_equ} is invariant under the sign change of $x$. Hence the right hand side must be even in $x$. That renders $u_0(x)$ and $u_1(x)$ either even or odd in $x$. For the Chalker Coddington network the critical point is at $x=0$. This lets us choose $u_0(x)$ odd and $u_1(x)$
even. The fit now should use as few coefficients as possible while reproducing the data as closely as possible.

The scaling function $\Gamma$ in the right side of \eqref{eq:ren_equ} is expanded in the fields $u_0$ and $u_1$ yielding
\begin{equation}
\label{expansin_in_fields}
\begin{split}
	\Gamma(&u_0(x)M^{1/\nu},u_1(x)M^y)= \Gamma_{00}+ \Gamma_{01} u_1M^y
	 +\Gamma_{20}u_0^2M^{2/\nu}\\	& + \Gamma_{02}u_1^2M^{2y}
	 +\Gamma_{21}u_0^2u_1M^{2/\nu}M^y +\Gamma_{03}u_1^3 M^{3y} \\
	& +\Gamma_{40}u_0^4M^{4/\nu}+\Gamma_{22}u_0^2 M^{2/\nu}u_1^2 M^{2y} + \Gamma_{04}u_1^4M^{4y}+\dots
\end{split}
\end{equation}
We further expand $u_0$ and $u_1$ in powers of $x$ as was done, for example, in Refs. \cite{SlevinOhtsuki2009, AmadoMalyshevSedrakyanEtAl2011}:
\begin{equation}
\label{fields_expanded}
 u_0(x)=x+\sum_{k=1}^\infty a_{2k+1}x^{2k+1} \quad \text{and} \quad
 u_1(x)=1+\sum_{k=1}^\infty b_{2k}x^{2k} .
\end{equation}
In Eq. \eqref{expansin_in_fields} we retained only terms that are even in $x$. Because of the ambiguity in the overall scaling of the fields, the
leading coefficient in Eq. \eqref{fields_expanded} can be chosen to be 1.

\subsection{Weights and Errors}\label{weight_error}

\begin{figure}[t]
  \beginpgfgraphicnamed{histogram-graphic}
  \begin{tikzpicture}
  	\begin{axis}[
  			width=\columnwidth,
  			height=6.5cm,
  			/pgf/number format/precision=3,
  			axis x line=bottom,
  			axis y line=left,
  			xlabel=$\gamma$,
  			ylabel=points per interval,
  			mark options={mark=none},
  			enlarge x limits=auto,
  		]
  
  		\addplot+[ybar, bar width=6.3pt, fill, opacity=0.5] coordinates{
        (0.015496, 2)
        (0.015515, 1)
        (0.015534, 4)
        (0.015552, 11)
        (0.015571, 10)
        (0.01559, 25)
        (0.015609, 40)
        (0.015628, 45)
        (0.015646, 62)
        (0.015665, 68)
        (0.015684, 68)
        (0.015703, 73)
        (0.015721, 60)
        (0.01574, 50)
        (0.015759, 35)
        (0.015778, 30)
        (0.015797, 23)
        (0.015815, 9)
        (0.015834, 10)
        (0.015853, 6)
      } 
      \closedcycle;
  
  		\addplot+[smooth] coordinates{
        (0.015496, 1.2368)
        (0.015515, 2.6405)
        (0.015534, 5.2108)
        (0.015552, 9.506)
        (0.015571, 16.0307)
        (0.01559, 24.9901)
        (0.015609, 36.0119)
        (0.015628, 47.9719)
        (0.015646, 59.0732)
        (0.015665, 67.2444)
        (0.015684, 70.7596)
        (0.015703, 68.8298)
        (0.015721, 61.8914)
        (0.01574, 51.4455)
        (0.015759, 39.5299)
        (0.015778, 28.0781)
        (0.015797, 18.4362)
        (0.015815, 11.1902)
        (0.015834, 6.2787)
        (0.015853, 3.2566)
      };
    \end{axis}
  \end{tikzpicture}
\endpgfgraphicnamed
	\caption{
	  Distribution of Lyapunov exponents in the ensemble of calculations
		with 624 elements for chain length $L=5\,000\,000$, \hbox{$M=60$} and
    \hbox{$x=0.02$}.
  }
	\label{fig1}
\end{figure}
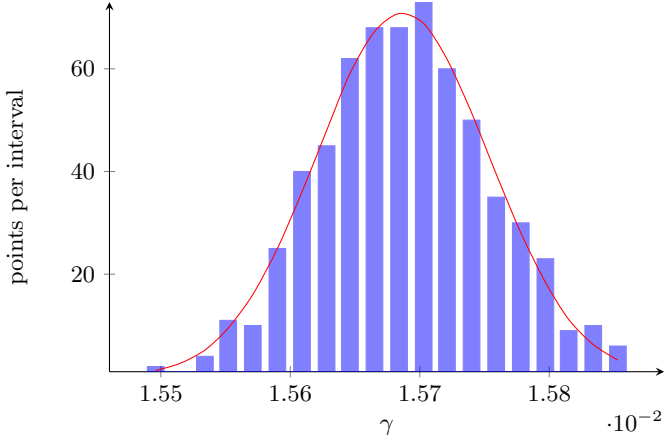

The left hand side of Eq.~\eqref{eq:ren_equ} is determined by the results of
numerical simulations of the random network model. Following
  Ref. \cite{AmadoMalyshevSedrakyanEtAl2011} we have produced large ensembles
  of the Lyapunov exponent $\gamma$ by simulating many disorder realizations
  for many combinations of $x$ and $M$. We calculated $624$
  disorder realizations for any combination of $M=20, 40, 60, 80, 100, 120,
  140, 160, 180, 200$ and $x=0.08/12\cdot [0, 1, 2, 3, 4, 5, 6, 7, 8, 9, 10, 11,
    12]$ for fixed $L=5\,000\,000$.  Our goal is to check whether the central
  limit theorem (CLT) \cite{tutubalin1965limit} also works in the case of
  randomness of the network or not. Fig.~\ref{fig1} shows the distribution of
  the Lyapunov exponent for $M=60$ and $x=0.02$ being nicely
  described by a Gaussian which demonstrates the validity of CLT.

 In the fitting procedure, the weight of each such $\gamma$ is given by the
 reciprocal of the variance of the corresponding ensemble. On the right hand
 side of Eq. \eqref{eq:ren_equ} the fitting formula depending on $x$ and $M$ is
 used. The coefficients of the expansion and the critical exponents are the
 fitting coefficients.

The fits are performed in several steps. First a weighted nonlinear least
square fit based on a trust region algorithm with specified regions for each
parameter is applied. The resulting parameters are used in a further weighted
nonlinear least square fit based on a Levenberg-Marquardt algorithm. Here no
limits are imposed on the fit parameters.  The last step is repeated until the
resulting parameters stop changing.

\subsection{Evaluation of fits}
The next step is the evaluation of the fit results. We present several methods
to do this.

Very common is the \emph{${\chi^2}$-test}. $\chi^2$ is given by
\begin{equation}
	\chi^2=\sum_i \frac{(y_i-f_i)^2}{\sigma^2}
\end{equation}
where $f_i$ is the value predicted by the fit and $y_i$ the measured
value. $\sigma$ is given by the standard deviation.
As our fit contains large ensembles of data points for the same $(x,M)$
coordinates, $\chi^2=0$ is not possible, actually it will be large due to the
huge number of data points. The way to deal with this behavior is to consider
the ratio $\chi^2$/\emph{degrees of freedom}. The expectation value for this
ratio is 1 for an ideal fit.  The \emph{degrees of freedom} is the number of
data points in the fit minus the number of fit parameters.

Deviations from 1 are evaluated by use of the cumulative
probability $P(\tilde\chi^2 < \chi^2)$ which is the probability of observing -- just for statistical reasons -- a sample statistic with a smaller $\chi^2$ value than in our fit. A small value of $P$, i.e.~a large value of the complement $Q:=1-P$ is taken as indicative for a good fit. However, values of $P$ lower than $1/2$ indicate problems in the estimation of the error bars of the individual data points.

Another criterion is based on the width of the
\emph{confidence intervals}. This quantifies the quality of the prediction
for a single parameter. We use 95\% confidence intervals
which means that for repeated independent generation of the same amount of
data and application of the same kind of data analysis the resulting
confidence intervals contain the true parameter values in 95\% of the
cases.

A most sensitive criterion is the \emph{Akaike information criterion} (AIC)
\cite{Akaike1974}. AIC is founded on information theory; Akaike found a formal
relationship between Kullback-Leibler information and likelihood theory.  This
finding makes it possible to combine estimation (i.e., maximum likelihood or
least squares) and model selection under a unified optimization framework.

Unlike in the case of hypothesis testing, AIC does not assume that the correct model is among the tested models. AIC rather offers a relative estimate of the information lost when a given model is used to represent the process that generates the data. This way, given a collection of models, AIC ranks those models if they are based on the same data. In this case a comparison to the best model can be calculated easily. In case a different data base has been used, the models cannot be ranked or compared.

For the calculations presented in this article we have been using the AICc, which is a small sample version of AIC or, more precisely, a second order bias correction.  AICc is also valid if $k$ is not small compared to $n$, where $n$ denotes the sample size and $k$ denotes the number of parameters, and is given by
\begin{equation}
	\text{AICc}=\text{AIC}+\frac{2k(k+1)}{n-k-1}.
\end{equation}
This formula holds exactly if the model is univariate, linear, and has
normally-distributed residuals, but may in other cases still be used unless a
more precise correction is known. Further details on the AIC and the AICc can
be found in \cite{BurnhamAnderson2002}.

The AIC can be expressed in terms of $\chi^2$:
\begin{equation}
	\text{AIC}=2k+\chi^2-2C
\end{equation}
Here $2C$ is a constant (dependent on the set of data points) that can be omitted because for comparisons we only need differences of AICc's.

For comparing models, the AIC (and the AICc) are used in the following way. Suppose, we have $l$ models with AIC$_1$, \dots AIC$_l$. The model with the smallest AICc --- let us call it AIC$_{min}$, --- is the favorite one. The relative probability of model $j$ compared to the model with  AIC$_{min}$ is
\begin{align}
\exp \frac{\text{AIC}_{min}-\text{AIC}_j}{2}.
\end{align}
Note that the exponential expression is smaller than one.

The last criterion we present is the sum of \emph{residuals}.
It is given by	$\mathit{res}=\sum_j \mathit{res}_j, \; \mathit{res}_j=y_j-f_j$. The sum of residuals should be small compared to the number of degrees of freedom. The residuals plotted should look like  noise around zero. If the residuals significantly deviate from zero, we expect that the fit function is not correct.

\subsection{Results}

In Fig.\ref{fig1} we present an example of the distribution of Lyapunov
exponents for fixed width $M$, parameter $x$ and chain length $L$. This
distribution defines the data point and its accuracy for the
  combination $(x,M)$. The reciprocal of the variance is used as the weight
  the data point carries in the fitting procedure.

\begin{figure}
  \input{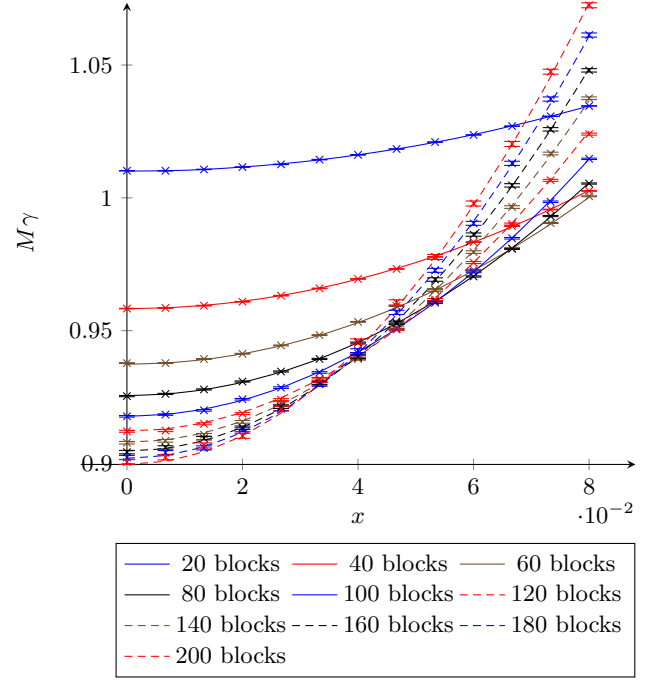}
	\caption{Plot of the smallest eigenvalue of the transfer matrix times
          $M$ (number of blocks) depending on the distance $x$ from the
          critical point. The $x$-values divide the interval $[0,0.08]$ into
          12 equal parts.}
	\label{fig2}
\end{figure}

In Fig.\ref{fig2} we present the product $M \gamma$ (the left-hand side of
Eq. (\ref{eq:ren_equ})) versus $x$ for various values of the width $M$. The
corresponding fitting parameters are presented in the table below.

Our best fitting results have been obtained by expanding $\Gamma$ up to second order in $u_0$ and $u_1$ \eqref{expansin_in_fields}, and expanding $u_0$ ($u_1$) up to the third (second) order in $x$. We found the following coefficients and goodness of fit parameters:
\\

\noindent
Coefficients (confidence bounds 95\%):
\begin{align*}
			 \hline \\[-2.5ex]
       \Gamma_{00} =\; &   \quad  0.864  & (0.856  &, 0.871)  \\
       \Gamma_{01} =\; &   \quad  0.0898 & (-0.071 &, 0.250)  \\
       \Gamma_{02} =\; &   \quad  0.976  & (0.907  &, 1.046)  \\
       \Gamma_{20} =\; &   \quad  0.312  & (0.302  &, 0.321)  \\
       a_3         =\; &   \quad  0.293  & (-0.221 &, 0.807)  \\
       b_2         =\; &   		 -0.255  & (-0.460  &,-0.049) \\
       \nu         =\; &   \quad  2.374  & (2.356   &, 2.391) \\
       y           =\; &         -0.356  & (-0.407  &,-0.306) \\
       \hline
\end{align*}
\noindent
Goodness of fit parameters:
\nopagebreak \vspace{-0.2ex}
\begin{align*}
  \hline \\[-2.5ex]
	& \chi^2\text{\,: } && 81192.5 \\
	& \text{degrees of freedom (dof) \,: } && 81112 \\
	& \chi^2/\text{dof\,: } && 1.001 \\
	& P\text{\,: } && 0.554 \\
	& \text{AICc\,: } && -556356.5 \\
	& \text{sum of residuals\,: }& & 181.86 \\
  \hline
\end{align*}

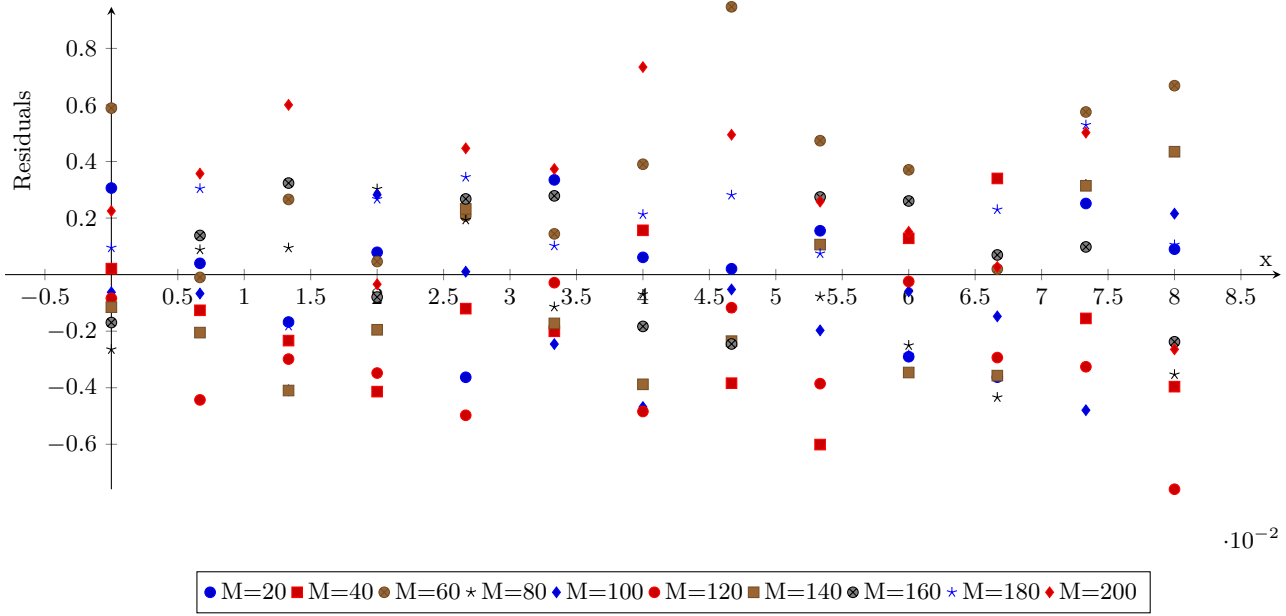
\begin{figure*}
  \begin{tikzpicture}
  \begin{axis}[
    width=\textwidth,
    height=8cm,
    /pgf/number format/precision=3,
    axis x line=middle,
    axis y line=middle,
    xlabel=x,
    ylabel=Residuals,
    ylabel style={at={(0,0.6)},rotate=90},
    enlarge x limits=auto,
    only marks,
    legend style={at={(0.15, -0.17)},anchor=north west},
    legend columns=10,
  ]
    \addplot table[x=x, y=M20]{table.csv};
    \addplot table[x=x, y=M40]{table.csv};
    \addplot table[x=x, y=M60]{table.csv};
    \addplot table[x=x, y=M80]{table.csv};
    \addplot table[x=x, y=M100]{table.csv};
    \addplot table[x=x, y=M120]{table.csv};
    \addplot table[x=x, y=M140]{table.csv};
    \addplot table[x=x, y=M160]{table.csv};
    \addplot table[x=x, y=M180]{table.csv};
    \addplot table[x=x, y=M200]{table.csv};

    \legend{M=20, M=40, M=60, M=80, M=100, M=120, M=140, M=160, M=180, M=200}

  \end{axis}
\end{tikzpicture}
  \caption{
    This figure presents a plot of the residuals. The $x$-axis shows the scaling
    parameter $x$ and the $y$-axis the residuals. For each pair $(x,M)$ the
    corresponding residuals are summed up and the result is shown in the plot.
    By inspection the x-axis is at the center of the scattered residuals. This
    indicates that there is no systematic deviation between the data points and
    the model equation.}
  \label{fig:residuals}
\end{figure*}

The degrees of freedom have been calculated from the number of data points
$624\cdot 13 \cdot 10$ minus 8, the number of fit parameters. We see
$\chi^2/$dof is close to 1 and the cumulative probability $P=0.554$ is close to
$1/2$, marking a good fit result. The sum of residuals is small compared to the
number of degrees of freedom. As can be seen in Fig.\ref{fig:residuals}, the
residuals are distributed around zero as judged by the eye. All this indicates
that the fit is reliable and the data agree with the model equation.

Fits with two irrelevant fields are clearly discouraged by the Akaike criterion.
Those models produce a (relative) Akaike coefficient of at least
$\text{AICc}=-556340$.  Therefore their relative likelihood is  about 0.0003.

\afterpage{\clearpage}



%

\end{document}